\documentclass[modern]{aastex63}

\usepackage{natbib,graphics,graphicx,mhchem,amssymb,lineno,commath}
\usepackage{appendix}
\bibpunct{(}{)}{;}{a}{,}{,}

\newcommand{\GG}[1]{}
\usepackage[symbol]{footmisc}

\begin{document}

\title{Io's SO$_2$ and NaCl Wind Fields From ALMA}

\correspondingauthor{Alexander E. Thelen}
\email{athelen@caltech.edu} 

\renewcommand{\thefootnote}{\fnsymbol{footnote}}
\setcounter{footnote}{1}

\author{Alexander E. Thelen}
\affiliation{Division of Geological and Planetary Sciences, California
  Institute of Technology, Pasadena, CA 91125, USA}

\author{Katherine de Kleer}
\affiliation{Division of Geological and Planetary Sciences, California
  Institute of Technology, Pasadena, CA 91125, USA}

\author{Martin A. Cordiner}
\affiliation{Solar System Exploration Division, NASA Goddard Space
  Flight Center, Greenbelt, MD 20771, USA}
\affiliation{Department of Physics, Catholic University of America, Washington,
  DC 20064, USA}

\author{Imke de Pater}
\affiliation{Department of Astronomy, Department of Earth and
  Planetary Science, University of California, Berkeley, CA 94720, USA}

\author{Arielle Moullet}
\affiliation{National Radio Astronomy Observatory, Charlottesville,
  VA 22903, USA}

\author{Statia Luszcz-Cook}
\affiliation{New York University, New York, NY 10003, USA}
\affiliation{American Museum of Natural History, New York, NY 10024, USA}

\begin{abstract}
We present spatially resolved measurements of SO$_2$ and
NaCl winds on Io at several unique points in its orbit: before and after
eclipse, and at maximum eastern and western elongation. The derived
wind fields represent a unique case of meteorology in a
rarified, volcanic atmosphere. Through the use of Doppler shift
measurements in emission spectra obtained with the Atacama
  Large Millimeter/submillimeter Array (ALMA) between $\sim346$--430
  GHz ($\sim0.70$--0.87 mm), line-of-sight winds up to $\sim-100$ m s$^{-1}$
in the approaching direction and \textgreater250 m s$^{-1}$ in the
receding direction were derived for SO$_2$ at altitudes of
  $\sim10$--50 km, while NaCl winds
consistently reached $\sim$\abs{150-200} m s$^{-1}$ in
localized regions up to $\sim30$ km above the surface. The
wind distributions measured at maximum east and west Jovian elongations, and on the
subJovian hemisphere pre- and post-eclipse, were found to be significantly
different and complex, corroborating the results of simulations that
include surface temperature and frost distribution, volcanic
activity, and interactions with the Jovian magnetosphere. Further, the
wind speeds of
SO$_2$ and NaCl are often inconsistent in direction and magnitude,
indicating that the processes that drive the winds for the two
molecular species are different and potentially uncoupled; while the
SO$_2$ wind field can be explained through a combination of
sublimation-driven winds, plasma torus interactions, and plume activity,
the NaCl winds appear to be primarily driven by the plasma torus. 
  
\end{abstract}

\section{Introduction} \label{sec:intro}
\renewcommand{\thefootnote}{\arabic{footnote}}
\setcounter{footnote}{0}

Io, the closest Galilean satellite to Jupiter, harbors a tenuous --
but dynamic -- atmosphere primarily composed of S-bearing products originating
from a combination of surface frost sublimation, volcanic
activity, and photodissociation. Atmospheric species can be circulated
longitudinally through a cycle of sublimation from and deposition into surface frost during the day
and night, respectively (see reviews in
\citealp{de_pater_21, de_pater_23}). Additionally, volcanic eruptions
can loft gases tens to hundreds of km above the surface at
significant (100s m s$^{-1}$) speeds. Neutral gas escape (e.g., Na, K) leads to the formation of
extended gas clouds co-orbiting with Io around Jupiter, while atomic and molecular
charge exchange with the Jovian magnetosphere produces ionized species
that are accelerated into Io's plasma torus and distributed throughout the rest of the
Jovian system (see \citealp{bagenal_23}, and references therein). Io's
ionosphere was discovered by the \textit{Pioneer
  10} spacecraft \citep{kliore_74}, while sulfur dioxide
(SO$_2$) -- the primary atmospheric constituent -- and volcanic
activity were found with
\textit{Voyager 1} \citep{hanel_79, pearl_79, morabito_79}.
The atmosphere and volcanic features have been subsequently monitored by \textit{Galileo},
\textit{Juno}, and a host of ground- and space-based observations
\citep{schneider_23}, including recent measurements with the James
Webb Space Telescope \citep{de_pater_23b}. Io's SO$_2$ atmosphere was predicted
to exhibit latitudinal and longitudinal variability due to the
sublimation of inhomogeneously distributed frost, redistribution through
pressure driven winds and volcanic outgassing, and convergence on the
terminator \citep{ingersoll_85, ingersoll_89, moreno_91,
  austin_00, moore_09}. Observations have shown that SO$_2$ is enhanced
by up to a factor of $\sim10$ on the antiJovian
hemisphere compared to the subJovian
hemisphere, and mainly confined to
$30$--$40^\circ$ above and below the equator \citep{lellouch_90, jessup_04, spencer_05, feaga_09, tsang_13a, jessup_15,
  lellouch_15, de_pater_20b,
  giono_21}, with evidence for temporal variability 
between perihelion and aphelion \citep{tsang_12, tsang_13b,
  giles_24}. Volcanic activity can help support the SO$_2$ atmosphere
(particularly when it condenses during eclipses by Jupiter),
and can also suffuse
additional sulfuric and alkali products -- such as NaCl
\citep{lellouch_03} and KCl
\citep{moullet_13} -- into the atmosphere. 

\begin{figure}
  \centering
  \includegraphics[width=0.95\textwidth]{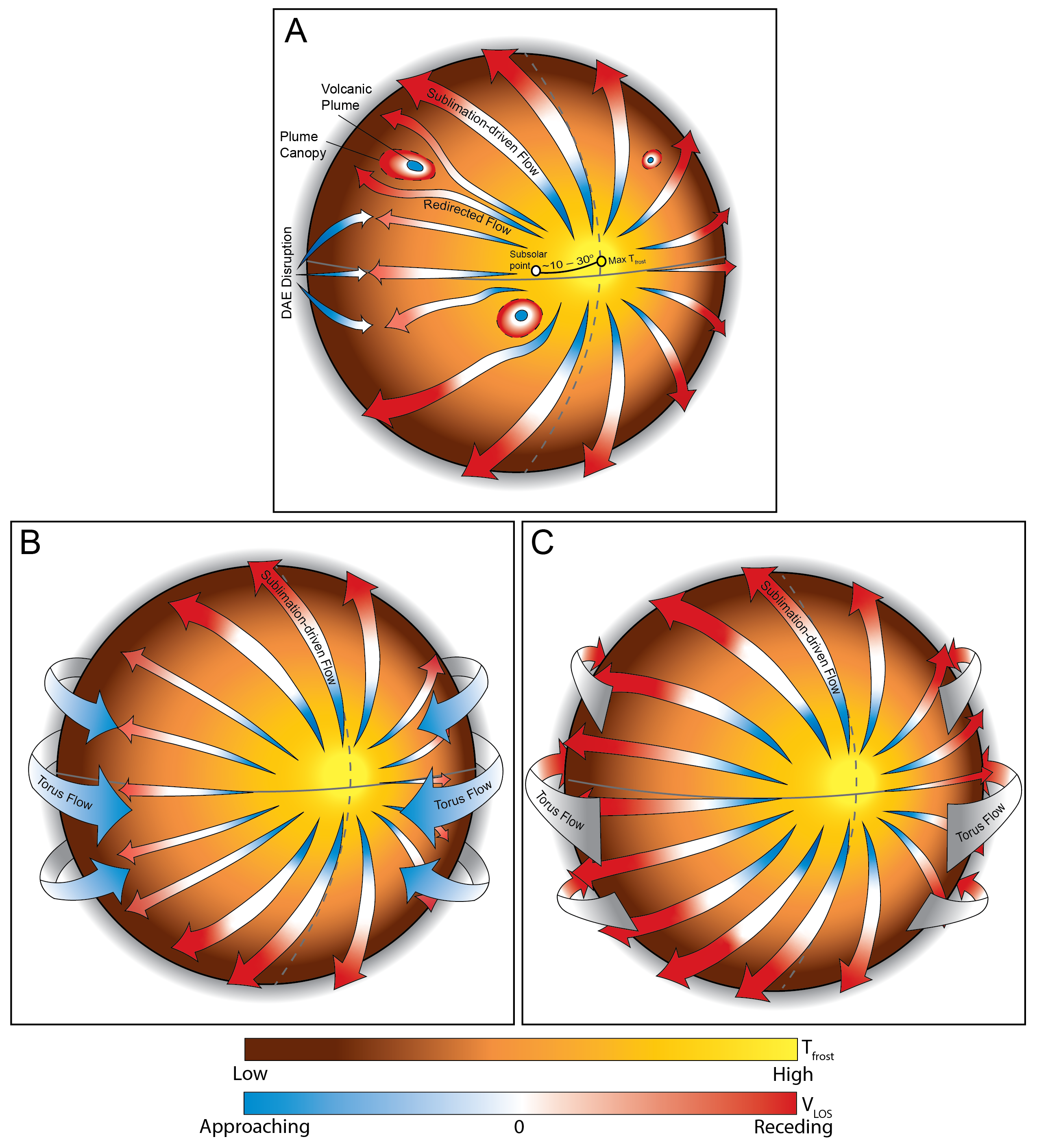}
  \caption{Diagram of Io's wind field as predicted by models and
    observable through line-of-sight (LOS) Doppler measurements. (A)
    Processes driving Io's wind field without the influence of the
  plasma torus, including sublimation-driven winds, volcanic plumes
  and surrounding canopies, and disruption of the day-to-night wind
  flow by the dawn atmospheric enhancement. Note that volcanic
  emission will be small compared to the ALMA beam, and the plume sizes are
  exaggerated for clarity here. (B) Schematic of Io's
  leading hemisphere showing the competition of
  sublimation-driven winds and the plasma torus, which impedes
  day-to-night flow at the dawn and dusk terminators. (C) Similar to
  B, but for Io's trailing hemisphere. Here, the plasma flow enhances
  the sublimation-driven flow at both terminators. In all panels, the
  orange color scale reflects the surface SO$_2$ frost temperature,
  which is offset from the sub-solar longitude by
  $\sim10-30^\circ$ (see models by, e.g., \citealt{walker_10}, and
  continuum observations by \citealt{de_pater_20b}). Blue gradients represent LOS winds moving
  radially towards the observer, while red represents LOS winds moving away
from the observer.}
  \label{fig:dia}
\end{figure}

While the distribution of Io's atmospheric species has been observed
and modeled throughout the previous decades \citep{de_pater_21, de_pater_23},
observations of winds on
Io remain sparse. Increasingly sophisticated simulations show the
expansion of sublimation-driven winds following atmospheric pressure gradients from the
surface location of peak frost temperature, which lags
$\sim10$--$30^\circ$ east of the sub-solar point, to Io's night side
\citep{walker_10, de_pater_20b}. These flows are complicated by the
inclusion of additional atmospheric species \citep{moore_09}, surface
thermophysical properties \citep{walker_12}, localized (and irregular)
volcanic outgassing \citep{mcdoniel_15, mcdoniel_17}, and upper atmospheric interactions with the
plasma torus \citep{walker_10, moore_11, walker_phd,
  mcdoniel_19}. These processes are depicted in Figure \ref{fig:dia}.

The resulting wind field may vary with atmospheric
species \citep{moore_11, walker_phd} -- some of which may even
manifest cyclonic winds -- and undoubtedly with volcanic
activity, which varies both spatially and temporally in an
unpredictable manner \citep{de_kleer_19c}. The predicted pressure
gradients across the disk should result in supersonic wind speeds (up to $\sim500$
m s$^{-1}$ at 10 km; \citealp{walker_12}), which may drive circulation
around most of the moon within Io's day ($\sim42$ hr); though yet to
be observed, an enhancement of SO$_2$ gas over the dawn
hemisphere (designated as the Dawn Atmospheric
Enhancement, or DAE, by \citealp{walker_10}) may disrupt the day-to-night
flow through the creation of a localized shock front (Figure \ref{fig:dia}A). Further, the SO$_2$ atmosphere 
collapses rapidly (on order minutes) during eclipse \citep{tsang_15,
  tsang_16, de_pater_20b}, possibly precluding a stable flow from being
established over complete diurnal timescales. The duration required to rebuild
the atmosphere following eclipse is $\sim10$ min, yet the time for winds
to achieve measurable speeds due to temperature gradients is predicted
to lag on order $\sim1$--2 
hr \citep{walker_12}. Thus, variation in wind speeds throughout Io's
day (particularly following
eclipses) may be observable. While planetary winds on the terrestrial and
Giant planets -- and Saturn's moon, Titan -- have been
inferred through cloud-tracking observations (see
review by \citealp{simon_22}, and references therein), stellar occultation
measurements (see review by \citealp{sicardy_23}), or through
atmospheric temperature fields (\textit{via} the thermal wind
equation, e.g., \citealp{flasar_05, achterberg_08}), none of these techniques
are applicable to the atmosphere of Io due to its rarefied nature. 

Observations of Io at (sub)millimeter wavelengths previously revealed
the distribution of SO$_2$, SO, and alkali
gases using a variety of single-dish and interferometric facilities \citep{lellouch_90, lellouch_92, lellouch_96,
  lellouch_03, moullet_10, moullet_13, roth_20, de_pater_20b,
  redwing_22, de_kleer_24}. \citet{moullet_08} made the first direct
measurements of atmospheric winds in Io's SO$_2$ using the
Institut de Radioastronomie
Millim{\'e}trique (IRAM) Plateau de Bure
Interferometer (PdBI). These observations showed prograde, super-rotating winds with a $330\pm100$ m s$^{-1}$ difference
from limb-to-limb on Io's leading hemisphere ($\sim90^\circ$W; eastern
Jovian elongation),
which were not predicted by prior models of day-to-night sublimation-driven
winds. Subsequent line-of-sight wind models by \citet{gratiy_10}
were able to produce a similar limb-to-limb wind magnitude; however, these
results showed large ($\sim300$ m s$^{-1}$) red-shifted velocities at
the dusk limb, and minor ($\sim50$ m s$^{-1}$) blue-shifted
velocities at the dawn limb, which were at odds with the
relatively evenly distributed red-shifted ($160\pm80$ m s$^{-1}$) and
blue-shifted ($170\pm80$ m s$^{-1}$) winds of the IRAM/PdBI
observations \citep{moullet_08}. Volcanic plume activity and plasma torus
momentum transfer could potentially drive the
observed apparent prograde flow. However, the low spatial resolution
($\sim0.5\times1.5"$) of the IRAM/PdBI observations provided
insufficient sampling of the wind field to distinguish between these
(and other) mechanisms implemented in wind models.

Recently, the Atacama Large Millimeter/submillimeter Array (ALMA) has
enabled higher angular resolution (\textless0.5$"$) observations in the
(sub)millimeter wavelength regime, allowing for
the derivation of zonal atmospheric winds on Titan \citep{lellouch_19,
  cordiner_20}, Jupiter \citep{cavalie_21}, Saturn \citep{benmahi_22},
and Neptune \citep{carrion_gonzalez_23}. Here,
we measure spatially resolved winds on Io at both maximum eastern and
western elongations, and shortly before and after eclipse, utilizing
ALMA observations previously obtained for the purposes of investigating 
atmospheric collapse \citep{de_pater_20b} and measuring isotope ratios
of S and Cl in molecular species \citep{de_kleer_24}. 

\section{Observations} \label{sec:obs}
The ALMA facility provides
exceptional angular resolution and spectroscopic capabilities through
the combination of 40--50 12-m antenna dishes, with receivers that can be tuned to
frequencies between 35--950 GHz ($\sim0.3$--8.6 mm) at spectral
resolutions down to 31 kHz. Together, these capacities facilitate the
derivation of Doppler shifts in myriad spectral lines, including the
sulfur-bearing \citep{de_pater_20b} and chlorinated \citep{redwing_22}
species -- and their isotopologues \citep{de_kleer_24} -- detected on Io
thus far. 

For this work, we have chosen three
ALMA datasets from ALMA Band 7 (275--373 GHz; $\sim0.8$--1.1 mm) and 8
(385--500 GHz; $\sim0.6$--0.8 mm) consisting of seven individual observations of Io in
sunlight; the observing parameters are detailed in Table
\ref{tab:obs}. Observations
of Io at both maximum Jovian elongations were acquired on 2021 July 15 and 23. These observations
were reduced in a similar manner to previous studies
\citep{de_pater_20b, de_kleer_24}, using the scripts provided by the ALMA Observatory in the
Common Astronomy Software Analysis (CASA) package \citep{mcmullin_07,
  casa_22}. For the purposes of this study, multiple ALMA integrations
on Io over a period of $\sim4.5$ hr were combined so as to improve the
spectral signal-to-noise ratio (SNR). The approximate sub-observer longitudes during the mid-points of
these observations were $\sim98^\circ$W and
$266^\circ$W for the leading and trailing hemisphere observations,
respectively, though the integration of multiple ALMA executions over
a $\sim4.5$ hr period resulted in some longitudinal smearing
($\sim38^\circ$). We leave the analysis of the individual, shorter
($\sim45$ min) ALMA scans from these data to future work.

Io's continuum emission was subtracted from the calibrated ALMA data using the CASA
\texttt{uvcontsub} task, which operates on the
complex visibility measurements obtained through the
Fourier transform of Io's brightness distribution on the sky as
observed by each antenna pair. Subsequent image deconvolution was performed using the CASA 
\texttt{tclean} algorithm \citep{hogbom_74} with pixel sizes =
$0.01"$. This included iterative phase self-calibration, as discussed
in \citet{de_pater_20b}. The observations were taken with 42 and 45 12-m antennas
spatially separated by up to $\sim2.5$ km, which resulted in images of Io at
a spatial resolution of
$\sim0.1''\times0.1''$ when using the Briggs weighting scheme
  \citep{briggs_95} during deconvolution (robust parameter = 0.5). We further convolved the native ALMA synthesized
beam shape (analogous to a point-spread function) from these
observations with a second Gaussian function using the
CASA \texttt{imsmooth} task to achieve a 
resolution of $0.20''\times0.20''$, so as to improve the spectral SNR
(and thus the measured velocity SNR) in
each pixel and to better facilitate the comparison to other observations.

The data from
observations in 2018 and 2022 were utilized directly from the previous analyses by
\citet{de_pater_20b} and \citet{de_kleer_24}, respectively. These
data include observations of Io at multiple times throughout its orbit:
shortly ($\lesssim30$ min) before (March) and after (September) eclipse in 2018, and at
both eastern and western Jovian elongations in 2022 May. While many SO$_2$
transitions, as well as those of SO, KCl, and
various isotopologues were detected in these observations, we only
utilize the strongest two to three SO$_2$ transitions, and the NaCl ($J=33-32$) transition from
the 2022 data, for this study (spectral transitions are tabulated in
Appendix \ref{app:maps}). Imaging of the ALMA
data from these projects with pixel sizes
of $0.04''$ (2018) and $0.03''$ (2022) produced image cubes with spatial scales of
$\sim0.22$--$0.35''$ when performing \texttt{tclean} with
  Briggs weighting (robust = 0.5), which were used again here
without further alteration. 

In all instances,
spectra were obtained with a native spectral resolution of 141 or 282 kHz
($\sim0.1$--0.2 km s$^{-1}$),
allowing for high precision velocity measurements when fitting a line
profile across multiple ($\sim9$--18) spectral channels. While additional
observations of Io during 2018 
exist, these were taken (intentionally) while Io was eclipsed by 
Jupiter \citep{de_pater_20b}. As such, the emission from sublimated SO$_2$ was
significantly reduced 
due to the rapid collapse of Io's atmosphere during eclipse (leaving
only weak emission from volcanic SO$_2$ gas), 
and did not permit the accurate measurement of winds. 

\begin{deluxetable*}{llllllll}
  \tablecaption{Observational Parameters \label{tab:obs}}
  \tablecolumns{11}
  \tablehead{Project & Date & Ang. & Ang. & Spec. Res. & Lat. & W. Lon. & Refs. \\
    Code $\#$ & & Dia.$^{a}$ ($''$) & Res. ($''$) & (kHz [km s$^{-1}$]) & ($^\circ$)
    & ($^\circ$) & \\ [-2.75ex]}
  \startdata

2017.1.00670.S & 2018 Mar. 20 & 1.06 & $0.35\times0.35$ & 141 [0.12] & -3.40
& 337.2 & [1] \\
& 2018 Sept. 02 & 0.89 & $0.30\times0.30$ & 141 [0.12] & -2.96 & 25.3 & [1] \\
& 2018 Sept. 11 & 0.87 & $0.22\times0.22$ & 141 [0.12] & -2.95 & 21.5 & [1] \\

2019.1.00216.S & 2021 July 15 & 1.20 & $0.20\times0.20$ & 141 [0.12] & 0.86 &
 98.0 &  \\
 & 2021 July 23 & 1.22 & $0.20\times0.20$ & 141 [0.12] & 0.86 & 266.0 & \\ 

2021.1.00849.S & 2022 May 18 & 0.92 & $0.30\times0.27$ & 282 [0.20] & 2.09 &
 277.1 & [2] \\
 & 2022 May 24 & 0.94 & $0.35\times0.23$ & 282 [0.20] & 2.14 & 76.8 & [2] \\

  \enddata
  \footnotesize
   \tablecomments{$^{a}$Column denotes the angular diameter of Io for
     each observation. References: [1] \citet{de_pater_20b}; [2]
     \citet{de_kleer_24}. Data from ALMA Project Code
     $\#$2019.1.00216.S has not been previously published.}
\end{deluxetable*}

\begin{figure}
  \centering
  \includegraphics[width=0.98\textwidth]{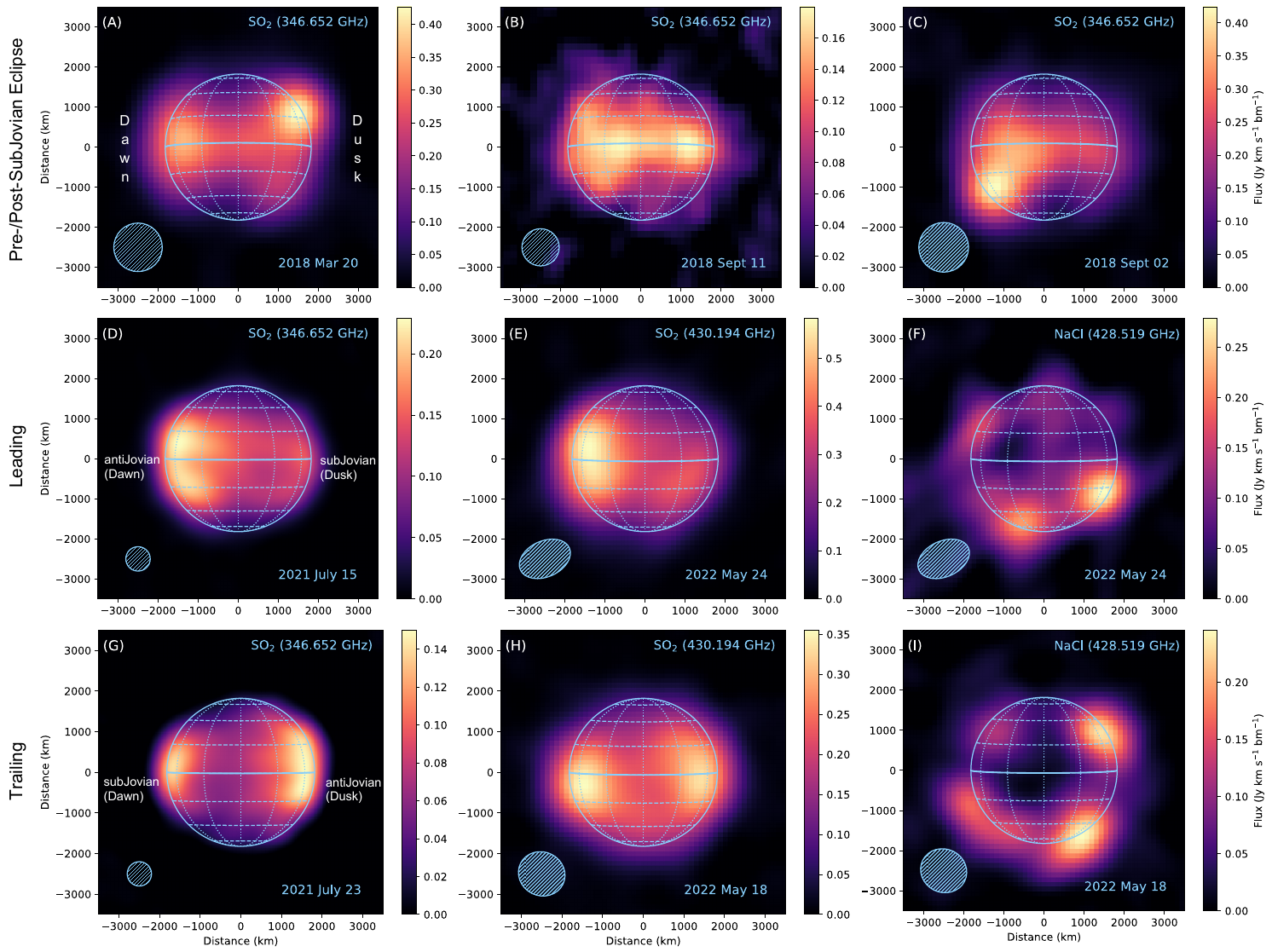}
  \caption{Integrated emission maps showing the
    distribution of Io's SO$_2$ as observed before (A) and following
    (B, C) eclipse, at maximum east (D, E) and west (G, H)
    elongations. The NaCl distributions observed in 2022 are shown in
    panels F and I. Io's dawn and dusk hemispheres, and the sub-
    and antiJovian directions (when relevant) are labeled for each row. Io's latitude and
    longitude grid are shown in blue in $22.5^\circ$ and $30^\circ$
    increments, respectively. The ALMA resolution element is shown in
    the bottom left as a hashed ellipse.}
  \label{fig:maps}
\end{figure}

The spatial 
distribution of Io's atmospheric SO$_2$ and NaCl is shown for each of these
observations in Figure \ref{fig:maps}, which was obtained by
integrating over the spectral line emission in each pixel. The
SO$_2$ and NaCl emission originates from the lower $\sim50$ km
of the atmosphere as depicted by the contribution functions in
\citealt{de_pater_20b} and \citealt{de_kleer_24} (see their Figures 12
and S3, respectively), which depend on the fractional coverage of the gas across the disk, the column density and temperature. As in prior observations of Io in the
(sub)millimeter \citep{moullet_10, de_pater_20b}, IR
\citep{spencer_05}, and UV \citep{feaga_09}, the SO$_2$ gas
distribution is mostly confined to low latitudes and enhanced on Io's antiJovian hemisphere at each maximum
elongation. A secondary, smaller enhancement is evident on the
trailing hemisphere (Figure \ref{fig:maps}G, H), which may be related
to the DAE proposed by \citet{walker_10}. The NaCl emission is highly localized, interpreted by
previous studies as a tracer of volcanic outgassing \citep{redwing_22, de_kleer_24}.

\section{Spectral Line Modeling $\&$ Fitting} \label{sec:mod}
\begin{figure}
  \centering
  \includegraphics[width=0.5\textwidth]{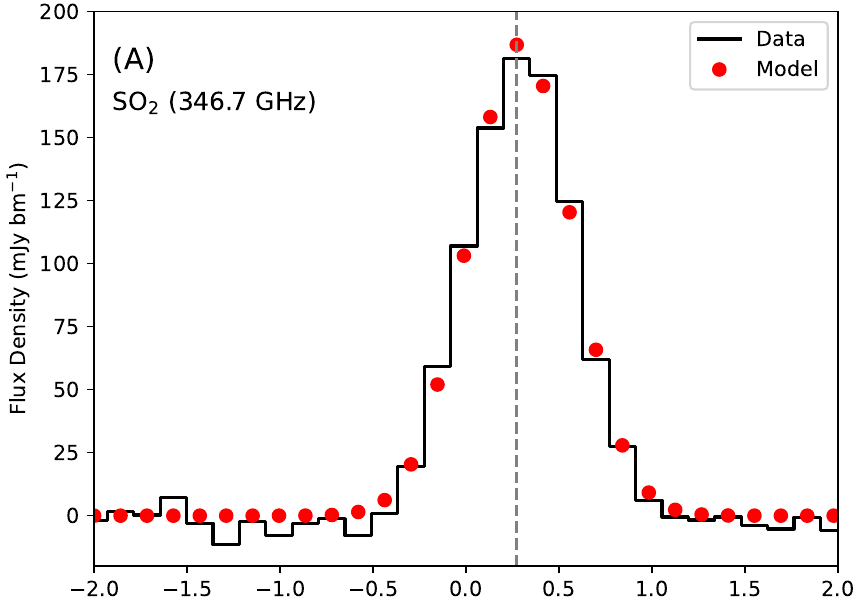}
  \includegraphics[width=0.5\textwidth]{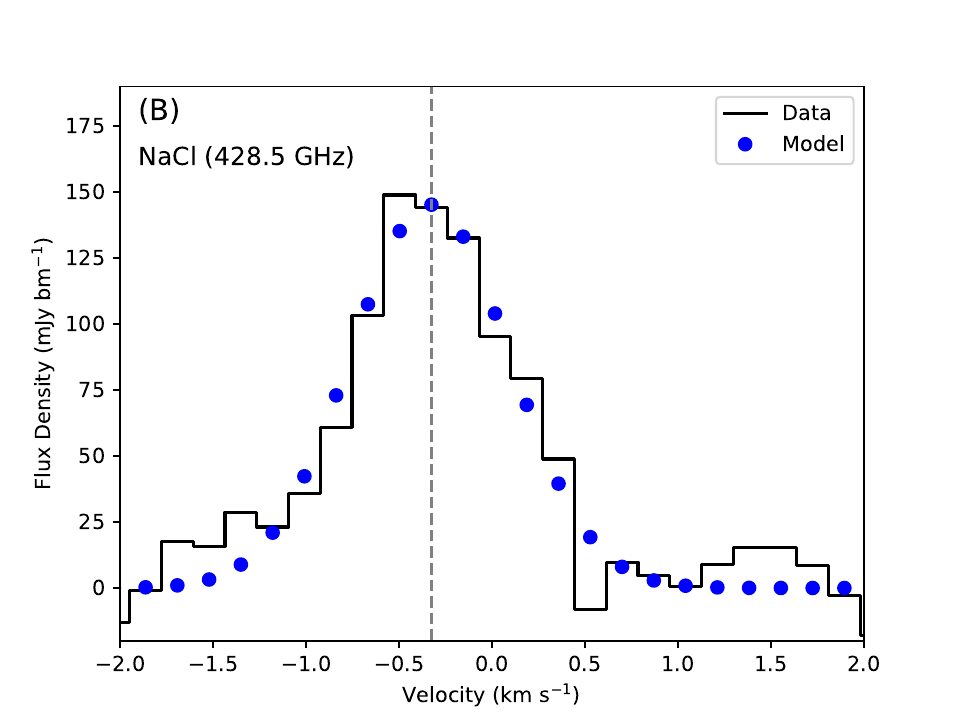}
  \caption{Spectra from individual pixels in ALMA images of Io (black)
    compared to best-fit model spectra (red, blue datapoints). The
    resulting LOS velocity fit is denoted by
    the dashed gray line. (A) SO$_2$ ($J_{K_a,K_c}=19_{(1,19)}-18_{(0,18)}$)
    transition at 346.7 GHz extracted from the equator on the dusk
    (antiJovian) limb of observations acquired on 2021
    July 23 targeting Io's trailing hemisphere (as shown in Figure
    \ref{fig:maps}G). The model fit
    results in a line-of-sight velocity of $259\pm11$ m s$^{-1}$. (B) NaCl
    ($J=33-32$) transition at 428.5 GHz extracted from the plume
    emission described by \citet{de_kleer_24} at $\sim22^\circ$N,
    $159^\circ$W (their `location 3') on 2022 May 24 (Figure \ref{fig:maps}F). A best-fit velocity of $-334\pm23$ m
    s$^{-1}$ was found.}
  \label{fig:spec}
\end{figure}

Following the acquisition of ALMA spectral line image cubes, the
LOS radial velocities across Io were derived by
fitting the spectral line shape from each pixel. These velocities were
then translated to LOS wind speeds by accounting for the moon's
solid-body rotation (75 m s$^{-1}$ at the limb), which can be 
substantial compared to the derived velocities, those previously
measured in the millimeter \citep{moullet_08}, and predicted by
models \citep{gratiy_10}. 
The methods used to derive Io's wind speeds (and associated
uncertainties) follow the procedures described in \citet{cordiner_20},
employed in that study for the measurement of zonal winds in Titan's
upper stratosphere, mesosphere, and thermosphere. Unlike
(sub)millimeter molecular
emission on Titan and the giant planets, Io's SO$_2$ and NaCl lines are not broadened by pressure
but only by Doppler thermal broadening and winds
\citep{lellouch_90, gratiy_10}. As such, these spectra were well-fit by a
Gaussian line shape. The spectrum extracted from each pixel was fit independently over a
$\sim6$--12 MHz (50 spectral channels) region
surrounding the spectral line. Starting with the gas temperature,
SO$_2$ and NaCl column densities found by \citet{de_kleer_24} for both
leading and trailing hemispheres of Io, a uniform slab model was used
to produce model spectra using a Gaussian line shape and varying the gas
column density. Here, we assumed local
thermodynamic equilibrium based on previously modeled rotational emission
in Io's lower atmosphere \citep{lellouch_92, gratiy_10,
  walker_10, de_pater_20b, de_kleer_24}. The \texttt{mpfit} module in
Python\footnote{https://cars9.uchicago.edu/software/python/mpfit.html}
was then used to derive the spectral line center of each pixel, the frequency of which
was compared to accurate rest frequency line centers from the Cologne
Database for Molecular Spectroscopy (CDMS\footnote{The CDMS can be
  accessed here: https://cdms.astro.uni-koeln.de/ ; see the SO$_2$ and
NaCl entries and references therein.}; \citealp{muller_01, muller_05, endres_16}) to
derive the LOS velocity. Example fits for SO$_2$
and NaCl spectra are shown in Figure \ref{fig:spec}, where LOS
velocities \textgreater\abs{250} m s$^{-1}$ were found; additional
information is detailed in Appendix \ref{app:maps}.

For each observation, the LOS velocity fields derived from two or three spectral
lines of SO$_2$ were combined into a weighted average. The
  different spectral transitions probe altitudes between $\sim10$--50
  km depending on the viewing geometry \citep{de_pater_20b,
    de_kleer_24}; however, the derived velocities were fairly
consistent across spectral transitions. Individual NaCl
  transitions are sensitive to altitudes below $\sim30$ km \citep{de_kleer_24}. We leave the investigation of
how the wind field varies as a function of altitude to future work, as
it requires new
observations with much higher spectral SNR over a smaller ALMA beam
footprint, allowing for more accurate retrieval of winds on
Io's limb where the altitude differences may manifest more readily.
The 1-$\sigma$ velocity errors for each pixel were derived using a Monte Carlo method by
fitting the derived Gaussian line shape from each spectrum to $\sim300$ synthetic spectra as
discussed in previous works \citep{lellouch_19, cordiner_20}. The wind
errors ($\Delta \nu_{wind}$) can also be estimated based on the spectral SNR and full-width
at half-maximum (FWHM) of
the emission line (often $\sim600$ kHz or $\sim0.5$ km s$^{-1}$; see \citealp{de_pater_20b,
  de_kleer_24}), the central line frequency ($\nu_0$; see Appendix \ref{app:maps}), the data channel width ($\Delta \nu$), and the speed
of light ($c$) 
as discussed in Appendix D of \citet{cavalie_21} using the empirical formula:
\begin{equation}
  \Delta \nu_{wind} \sim \frac{FWHM}{SNR \times \sqrt{FWHM/\Delta
      \nu}} \times \frac{c}{\nu_0}
\end{equation}
For the 2021 data,
we find $\Delta \nu_{wind} \sim8$ m s$^{-1}$ using this formula for spectra
with SNR\textgreater10; these errors are
commensurate with the errors derived through the Monte Carlo analysis. To
assist in interpretation of the derived velocity and wind maps, pixels
with distances away from Io's surface greater than the ALMA beam
half-width ($\sim600$--1200 km) were masked out, as well as pixels with spectra
SNR\textless10. The errors associated with these pixels are often
$\gtrsim25$ m s$^{-1}$, approaching the magnitude of velocity values
found on the disk. The weighted average LOS velocity maps and the
associated maps of velocity uncertainty
are shown and discussed in Appendix \ref{app:maps}. 

Finally, Io's solid-body rotation at every latitude and longitude was calculated and
subtracted from the LOS velocity field. The solid-body rotation model
was convolved with the appropriate ALMA beam size and shape before
subtraction to account for smearing of the velocities across the
beam. A comparison of the data and rotation models is shown in
Appendix \ref{app:rot}. With the solid-body rotation subtracted, the
derived velocities now represent the LOS winds as observed at each
pixel location. 

\section{Results $\&$ Discussion} \label{sec:dis}
\begin{figure}
  \centering
  \includegraphics[width=0.98\textwidth]{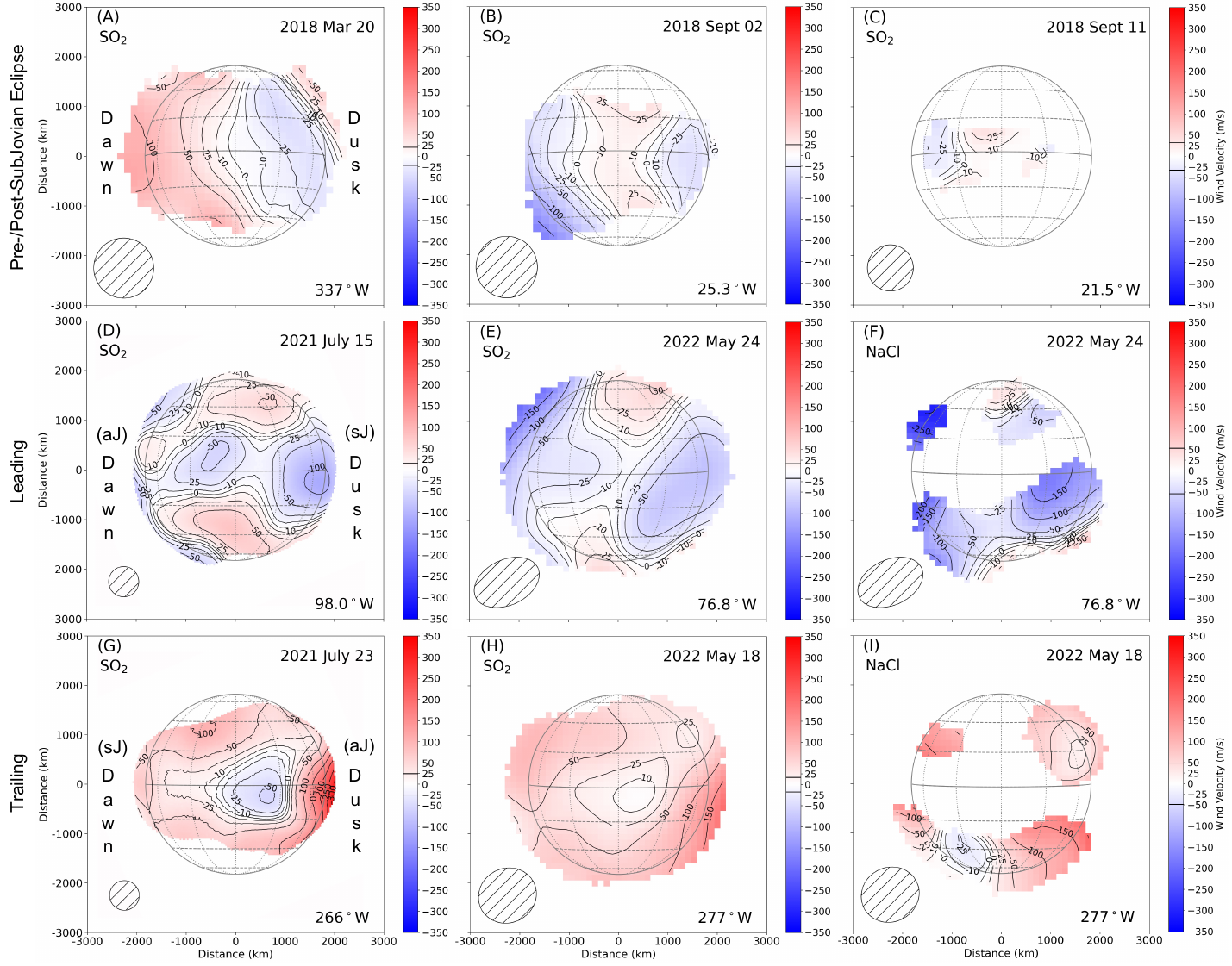}
  \caption{Derived line-of-sight wind maps following the Doppler shift fitting of
    Io's velocity field and the subtraction of Io's solid-body
    rotation. Only pixels with spectral
    SNR $\geq10$ are shown. Blue velocities correspond
    to winds approaching the observer, while red velocities are receding. The corresponding velocity
    maps and uncertainties are shown in Appendix \ref{app:maps}. Wind
    maps are grouped loosely by observation and
    longitude. Observational parameters are detailed in Table
    \ref{tab:obs}. Panels A--C show SO$_2$ winds on the subJovian
    hemisphere as Io entered (A) and exited (B, C)
    eclipse in 2018; 
    panels D and G show SO$_2$ winds at
    Jupiter's eastern and western elongation, respectively, from 2021;
    panels E, F, H, and I show SO$_2$ and NaCl winds at both Jovian
    elongations from 2022. 
    Io's dawn and dusk hemispheres are
    labeled in each row, as are the sub- and antiJovian directions
    (sJ and aJ, respectively). Io's solid-body radius,
    latitude and longitude lines are shown as a gray grid -- latitude
    and longitude contours increase in increments of $22.5^\circ$ and
    $30^\circ$, respectively. The central sub-observer longitude is
    denoted in the lower right of each panel. The corresponding ALMA beam (FWHM of the
    PSF) for each observation is shown
  as a hashed ellipse in the bottom left of each panel. The median 3-$\sigma$
wind errors are shown as solid lines on the color bar for each panel
for reference;
see Appendix \ref{app:maps} for the full uncertainty maps.}
  \label{fig:winds}
\end{figure}

The full set of derived wind maps is shown in Figure \ref{fig:winds}
following the subtraction of Io's solid-body rotation,
with positive (red) pixels representing winds moving away from the
observer, and negative (blue) pixels for winds moving towards. For
reference, the median 3-$\sigma$ velocity errors are denoted on the
colorbar for each map (black lines), allowing for the assessment of the significance
of the wind field in each location; the full error maps are shown in
Appendix \ref{app:maps}. It should also be noted that though the
atmosphere may experience (potentially strong) winds at all latitudes
and longitudes, the spatial distribution and SNR of the
SO$_2$ and NaCl lines measured in these observations constrained
our ability to measure the wind field at all locations (see
Figure \ref{fig:maps}). Additionally,
the Doppler winds were derived from radial velocities
and thus only represent the LOS component.

\subsection{Eclipse Observations} \label{sec:eclip}
Figure \ref{fig:winds}A--C shows the distribution of SO$_2$
winds from Io's ingress into (panel A) and egress from (panels B, C) eclipse by
Jupiter. While the observations were originally designed to measure
the distribution of SO$_2$ pre- and post-eclipse on the subJovian
hemisphere, they were utilized here to place constraints on two aspects of the wind
field: how long the wind field takes to manifest after Io experiences
daylight following eclipse, and to what degree pressure from the
plasma torus impacts the measured SO$_2$ winds, since they provide
measurements of Io when the local plasma velocity has no LOS component. 

Figure \ref{fig:winds}A shows the distribution of SO$_2$ winds on Io's subJovian hemisphere
$\sim30$ min before entering eclipse on 2018 March 20. Here, we find
receding winds at most longitudes on Io's dawn hemisphere, reaching 
speeds \textgreater100 m s$^{-1}$ towards the equatorial
limb. Winds approaching the observer are detected
towards the dusk limb at reduced magnitudes, though at the
$\sim3$-$\sigma$ level. The impact of the plasma torus on
Io's wind field is predicted to be complex \citep{walker_phd}, but viewing
Io's subJovian hemisphere removes the majority of the perpendicular
wind component due to the torus flow; as
such, Figure \ref{fig:winds}A can be used as a proxy for the underlying wind field before Io goes into
eclipse, which may be indicative of the subJovian winds caused mainly
by sublimation-driven gradients (see Figure \ref{fig:dia}A). Indeed, the wind field predicted by
models using the SO$_2$ frost and temperature distributions on Io
\citep{gratiy_10, walker_12} show winds moving in day-to-night streams
from equatorial
latitudes (seen as receding winds in our LOS observations) and lagging
$\sim30^\circ$ east of the sub-solar
longitude, where the peak surface frost temperatures result in slower
($\sim-10$--100 m s$^{-1}$) winds approaching the observer, similar to
what we measure here.

Winds from an observation $\sim30$ minutes following eclipse is shown in Figure
\ref{fig:winds}B from 2018 September 02. Here, we measure 
approaching winds at Io's dawn limb reaching maximum speeds around $-100$
m s$^{-1}$. Winds at the dusk limb are also approaching, though at
a reduced speed, and winds up to $\sim25$ m s$^{-1}$ (at the
$3$-$\sigma$ level) are receding at the central longitude.
Figure \ref{fig:winds}C shows the wind field from a second observation
following eclipse, which started just
after ($\sim7$ min) Io's egress from behind Jupiter's shadow on 2018
September 11. As shown
in \citet{de_pater_20b}, the emission of SO$_2$ at this time was faint, and
confined to low latitudes. The wind field we measure here is
statistically insignificant (i.e., the winds are \textless3-$\sigma$)
in all pixels. The comparison of the SO$_2$ distributions in Figure
\ref{fig:maps}B and C shows that the SO$_2$ atmosphere has rebuilt in
\textless30 min following eclipse \citep{de_pater_20b}; similarly, the
wind fields in Figure \ref{fig:winds}B and C shows the rapid build up
of winds during this period. The
sublimation-driven winds were predicted to be weak for $\sim1$ hr following egress
from eclipse and may originate $\sim10^\circ$E of the subsolar point
by \citet{walker_12},
which may explain the low-magnitude receding winds observed in Figure
\ref{fig:winds}B. The higher magnitude approaching winds along the
dawn hemisphere may be due to the emerging dawn atmospheric
enhancement, which takes
$\sim40$ min to establish following eclipse \citep{walker_12}, and may
dominate the sublimation-driven winds at this point following eclipse.

\subsection{Leading Hemisphere Winds} \label{sec:lead}
The maps in Figure \ref{fig:winds}D--F show the SO$_2$ and NaCl winds
measured on Io's leading hemisphere ($\sim90^\circ$W longitude, at
maximum eastern Jovian elongation)
during 2021 and 2022. At this point in its orbit, Io has experienced
significantly more daylight than the post-eclipse observations (Figure
\ref{fig:winds}B and C), so the SO$_2$ emission (Figure
\ref{fig:maps}D, E) was strong enough to
measure the winds at most latitudes and with smaller errors. Though
the SO$_2$ observations (Figure \ref{fig:winds}D,
E) are temporally separated
and are at different spatial resolutions (denoted by the ALMA beam in
the lower left of each panel), the wind distributions are largely
consistent. We observe approaching winds at speeds ranging from
roughly $-10$ to $-100$ m
s$^{-1}$ along the equator and towards the limbs, with receding winds
up to 50 m s$^{-1}$ towards higher latitudes, which are greater in
magnitude than those
tentatively measured post-eclipse in Figure \ref{fig:winds}B. These receding winds
are again likely tracers of the day-to-night flow driven by equatorial
sublimation-driven pressure
gradients, while the sublimation itself may manifest as winds in the approaching
direction along the equator (Figure \ref{fig:dia}A). The
wind field we observe on the leading hemisphere is inconsistent with that measured by
\citet{moullet_08} with the IRAM/PdBI, where receding winds up to
$\sim150$ m s$^{-1}$ were observed on the dusk hemisphere. A
significant difference ($\sim$\abs{100-150} m s$^{-1}$) in the 
wind flow is required to reconcile the winds measured between the ALMA
and PdBI observations on the leading hemisphere dusk limb, as the wind field we find
in both observations does not resemble a prograde flow. 

The Jovian plasma torus is impacting the opposite
hemisphere to those observed in Figure \ref{fig:winds}D--F, and thus may reinforce
the approaching wind flow as it wraps around the moon and transfers momentum to the atmosphere (see Figure \ref{fig:dia}B). In the SO$_2$ wind field, the approaching winds are not
significantly increased in magnitude compared to the observations of
Io following eclipse egress (Figure \ref{fig:winds}B), where the toroidal motion was
roughly perpendicular to the line of sight. Increased winds at the dusk limb
(particularly in Figure \ref{fig:winds}D) may be evidence of torus momentum transfer,
as the winds are predicted to be largely receding towards the
dusk terminator \citep{gratiy_10, walker_12,
  walker_phd}; receding winds on the dusk hemisphere were also observed by \citet{moullet_08}. The winds measured using NaCl emission (Figure
\ref{fig:winds}F) present additional evidence for the large influence of
upstream torus pressure. We measure only approaching NaCl winds (i.e., in the
direction of Io's orbit and the rotation of the plasma torus), reaching velocities up to $-300$ m
s$^{-1}$. No evidence of the day-to-night, sublimation-driven flow is observed in
the NaCl wind field due to the lack of measurable receding winds;
however, this may also be the result of the localized nature of the
NaCl emission. 

\subsection{Trailing Hemisphere Winds} \label{sec:trail}
The winds measured from SO$_2$ and NaCl emission lines on Io's
trailing hemisphere ($\sim270^\circ$W longitude, western Jovian
elongation) in 2021 and 2022 are shown in Figure \ref{fig:winds}G--I. The trailing hemisphere winds are
significantly different from those observed on the leading
hemisphere. Here, we observe primarily receding winds, which reach
velocities \textgreater200 m s$^{-1}$ away from the observer at Io's
dusk limb. At the higher resolution observation (Figure \ref{fig:winds}G),
we also measure winds approaching the observer at speeds reaching
roughly $-50$ m
s$^{-1}$ localized along the equator. This observation represents the best comparison to models of
the Ionian wind field, which predict winds emanating from along the
equator lagging the sub-solar longitude and receding winds at other
latitudes and longitudes following the day-to-night streams (Figure \ref{fig:dia}A). The flow
at the dawn limb is counteracted by Io's rotation and
potentially through the SO$_2$ DAE predicted by \citet{walker_10},
which would deplete (and possibly collapse) the westward flow towards the terminator. While
the DAE in SO$_2$ emission has yet to be definitively observed \citep{de_pater_23}, an enhancement of
SO$_2$ is apparent on the subJovian hemisphere in our trailing hemisphere observations from 2021 and 2022  (Figure
\ref{fig:maps}G and H, respectively). A pressure differential between the high column
density SO$_2$ region on the limb and longitudes just experiencing
sunlight to the west may interfere with the sublimation-driven flow
\citep{walker_12}. We do not observe the approaching wind feature near
the sub-observer longitude in Figure \ref{fig:winds}H, though the relatively larger
ALMA resolution here may smear the localized approaching winds across
the beam shape, resulting in
decreased receding winds at similar longitudes along the
equator. Slight differences in SO$_2$ frost
distribution, or temporal variability in volcanic activity, could also
decrease the approaching winds measured between the two
observations.

As the toroidal momentum experienced by Io's trailing
hemisphere would reinforce receding winds (i.e., the sub-solar
and sub-plasma points are approximately equal here), the increased speeds at
Io's dusk limb compared to those post-eclipse (Figure
\ref{fig:winds}A) may be indicators of this effect (see the diagram in
Figure \ref{fig:dia}C); compared to the
dusk limb of the pre-eclipse observation, the winds
measured at maximum western elongation are significantly ($\sim200$ m s$^{-1}$)
higher. On the contrary, the dawn winds exhibited by the 
observations on the trailing hemisphere are somewhat diminished compared to the
pre-eclipse elongation ($\sim50$ m s$^{-1}$ and $\sim100$ m s$^{-1}$, respectively)
which confounds this picture. Invoking the DAE would help to explain the
discrepancy between the dawn and dusk limbs, as both would
experience momentum from the plasma torus but only winds on the dawn
terminator would be disrupted by counter (approaching) flow from the
DAE. The consistent 50--200 m s$^{-1}$ winds
we observe in the trailing hemisphere NaCl map (Figure \ref{fig:winds}I) lends
credence to the notion that the NaCl winds are primarily driven by
torus interactions, as on the leading hemisphere (Figure \ref{fig:winds}F) -- at the
viewing geometry of this observation, NaCl LOS velocities driven purely by
volcanic activity seems unlikely.

\section{Conclusions: An Emerging Picture of Io's Wind Field} \label{sec:conc}
Through the analysis of LOS Doppler-shift maps from multiple ALMA
observations of Io, we have found
striking differences in Io's wind distributions as a function of
longitude and differences in wind speed magnitude with molecular
species. Unlike previous observations of the prograde, super-rotating wind
field by \citet{moullet_08}, the winds we measure here do not resemble
those found on other planetary bodies measured thus far, likely due to
the combination of low atmospheric pressures, volcanic and plasma phenomena
present on Io that impact the atmospheric flow. By assessing the maps
from SO$_2$ and NaCl at different viewing
geometries, we can draw some broad conclusions. Winds in the bulk
SO$_2$ atmosphere within $\sim30^\circ$ of the subsolar point are
receding at $\sim25$-50 m s$^{-1}$ on both hemispheres, driven by
day-to-night pressure and temperature
gradients. Localized approaching winds are detected in variable
locations along the equator, and can be explained by SO$_2$ frost
sublimation from a heterogeneous SO$_2$ frost distribution. Finally,
the largest magnitude winds we observe in SO$_2$ at each maximum elongation are in the
direction of Io's orbital motion, and also the direction of the rapid
plasma torus rotation, which overtakes Io at a relative velocity of 57 km s$^{-1}$. This, combined with the
NaCl wind maps that only show motion in the orbital direction, is evidence
that torus interactions -- such as momentum transfer -- alter the SO$_2$ wind field, and may be the
major drivers of the NaCl winds. However, the complex SO$_2$ wind
fields and the number of processes contributing to the SO$_2$ winds
make it difficult to determine the relative contribution of torus
pressure compared to other proposed mechanisms. 

It is clear that Io's wind field at low altitudes ($\lesssim50$ km) is influenced by both external (the
Jovian plasma
torus) and Iogenic (frost sublimation, volcanic plume) activity, and
is both locally and temporally variable. Further, our ability to probe
the winds depends on the availability of strong atmospheric rotational
line emission,
which can be asymmetrically distributed across Io. In the future, it
will be important to
analyze the wind field derived from multiple atmospheric species at
multiple times during Io's orbit, and potentially at multiple times
throughout its year. Further observations at higher angular
resolution, across multiple sub-observer longitudes and 
atmospheric species, would help to elucidate the connection between
these processes in a more systematic way. If taken concurrent with
observations that trace volcanic thermal emission, future
measurements could reinforce the somewhat tenuous connections we
observe here in the processes on the surface and in the Jovian 
environment that drive Io's complex meteorology.

\section{Acknowledgments}
Funding for this work was provided by the NASA ROSES Solar System
Observations program for A.E.T. and M.A.C. K.d.K. acknowledges support from the National Science Foundation (NSF)
under Grant $\#$2238344 through the Faculty Early Career Development
Program. I.d.P. and S.L-C. acknowledge funding from NASA Solar System
Workings Grant $\#$80NSSC24K0306, as a subawardee of the lead campus, University of Texas at Austin.

This paper makes use of the following ALMA data:
ADS/JAO.ALMA$\#$2017.1.00670.S, 2019.1.00216.S, and
2021.1.00849.S. ALMA is a partnership of ESO (representing its member
states), NSF (USA) and NINS (Japan), together with NRC (Canada), MOST
and ASIAA (Taiwan), and KASI (Republic of Korea), in cooperation with
the Republic of Chile. The Joint ALMA Observatory is operated by ESO,
AUI/NRAO and NAOJ. The National Radio Astronomy Observatory is a facility of the National Science Foundation operated under cooperative agreement by Associated Universities, Inc.

\begin{appendices}
  \section{Velocity Fitting and Maps} \label{app:maps}
\begin{figure}
  \centering
  \includegraphics[scale=0.6]{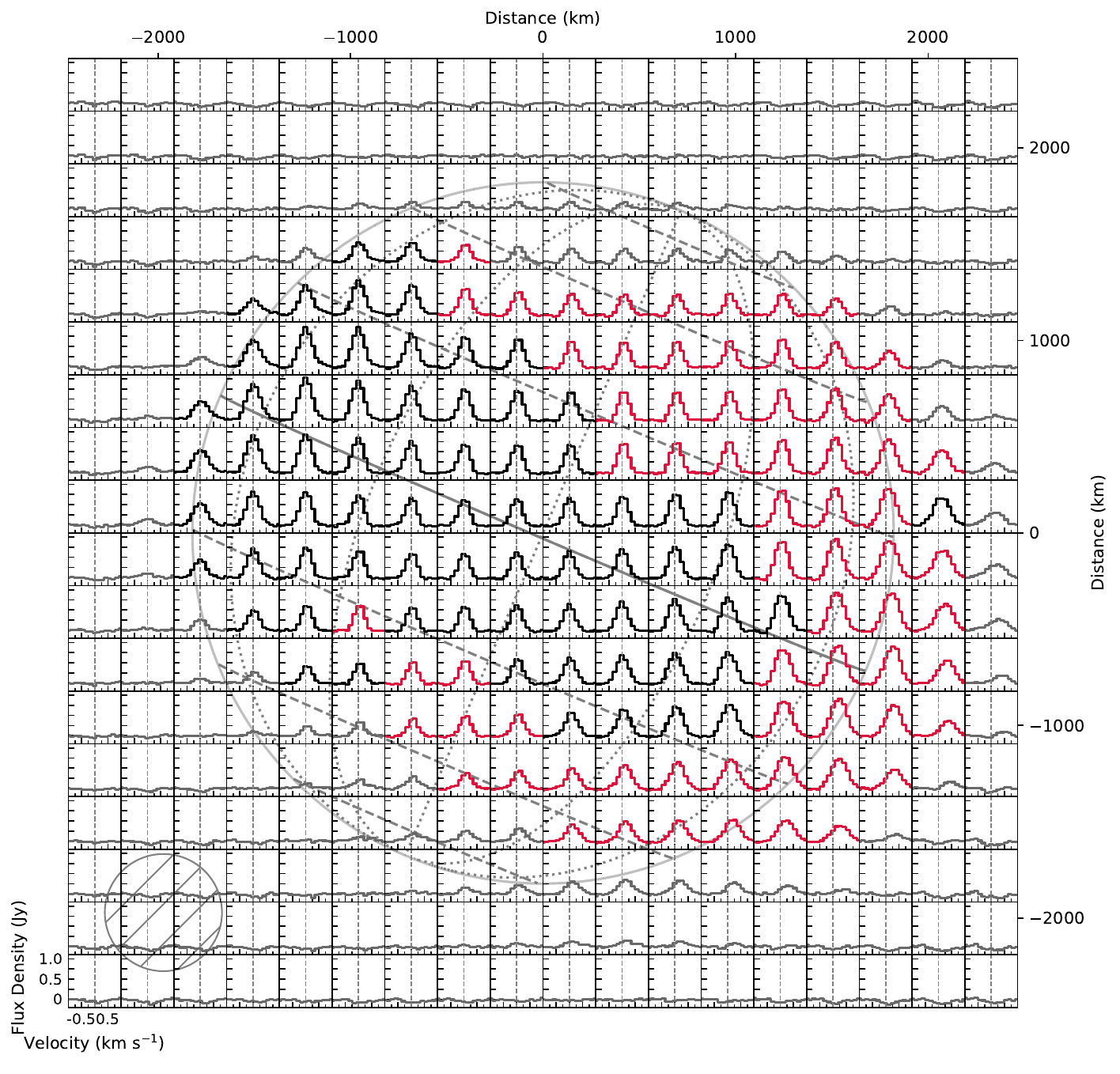}
  \caption{Grid of continuum-subtracted SO$_2$ ($J_{K_a,K_c}=19_{(1,19)}-18_{(0,18)}$) spectra from
    observations of Io's trailing hemisphere from 2021 July
    23. Spectra are presented in units of velocity; vertical dashed
    grey lines indicate the rest frequency of the line (zero velocity
    shift). The spectrum in each panel corresponds to the average of 100 pixels
    contained within each $10\times10$ pixel region. The flux density and velocity scales are shown on the lower
    left panel. Spectra
    with SNR\textless10 are shown in gray, and those with
    SNR $\geq10$ are shown in black. To demonstrate the
    observed Doppler shifts, spectra are colored red 
    where the frequency of the maximum channel falls 
    to the right of the SO$_2$ rest frequency by \textgreater1/2 of a
    spectral channel (61 kHz; $\sim60$ m s$^{-1}$). Io's solid-body radius is shown, as are latitude and
    longitude lines at 22.5$^\circ$ and 30$^\circ$ increments,
    respectively, in gray. Note that the north pole of Io is not
    rotated to align with the y-axis. The ALMA resolution element is shown as the
    gray hashed
    ellipse in the bottom left, which smears emission off of Io's disk.}
  \label{fig:spec_map}
\end{figure}

An example of spectral Doppler shifts exhibited by the Io data are
shown in Figure \ref{fig:spec_map}, which displays averaged SO$_2$ spectra
extracted from $10\times10$ pixel areas originating from the location
of each box with respect to Io's surface (gray grid in the background). For
averaged spectra with sufficient SNR, the
location of the SO$_2$ line center can be indicative of the Doppler
shift produced through the radial velocity 
induced by Io's rotation and wind field in the approaching (blue-shift) or receding (red-shift) directions. The derivation of Io's
winds is made more accurate by fitting the spectrum from each pixel
with a Gaussian function, as described in Section \ref{sec:mod}. The retrieved
central frequencies are then compared to those found by laboratory
measurements and previous observations, such as those documented on
the CDMS; the full spectral line parameters for the SO$_2$ and NaCl
transitions analyzed here are listed in Table \ref{tab:obs_spec}.

\begin{deluxetable*}{lllll}
   \tablecaption{Observed Spectral Transitions}
   \tablecolumns{11}
   \tablehead{Project & Species & Transition$^a$ & Rest Freq.$^b$ & $E_u$ \\
  Code $\#$ & & ($J_{K_a,K_c}$) & (MHz) & (K) \\ [-2.75ex]}
  \startdata

2017.1.00670.S  & SO$_2$ & $16_{(4,12)}-16_{(3,13)}$ & 346523.8784 & 164.5 \\
& SO$_2$ & $19_{(1,19)}-18_{(0,18)}$ & 346652.1691 & 168.1 \\

2019.1.00216.S & SO$_2$ & $19_{(1,19)}-18_{(0,18)}$ & 346652.1691 & 168.1 \\
& SO$_2$ & $20_{(0,20)}-19_{(1,19)}$ & 358215.6327 & 185.3 \\ 

2021.1.00849.S & SO$_2$ & 28$_{(5,23)}-28_{(4,24)}$ & 416825.5576 & 435.9 \\
 & SO$_2$ & $31_{(4,28)}-31_{(3,29)}$ & 419019.0378 & 497.0 \\
 & SO$_2$ & 24$_{(1,23)}-23_{(2,22)}$ & 430193.7070 & 280.5\\
 & NaCl & $33-32$ & 428518.5512 & 350.1 \\
  \enddata
   \footnotesize
   \tablecomments{$^a$Spectral transitions are written from
     the rotational upper ($J_{u}$) to lower ($J_{l}$) energy states
     with the projections onto the A and C inertial axes as $_{K_a}$
     and $_{K_c}$, respectively. $^b$Rest frequencies are extracted
     from the CDMS entries for SO$_2$ and NaCl from the references
     therein.}
   \label{tab:obs_spec}
 \end{deluxetable*}

\begin{figure}
  \centering
  \includegraphics[width=0.98\textwidth]{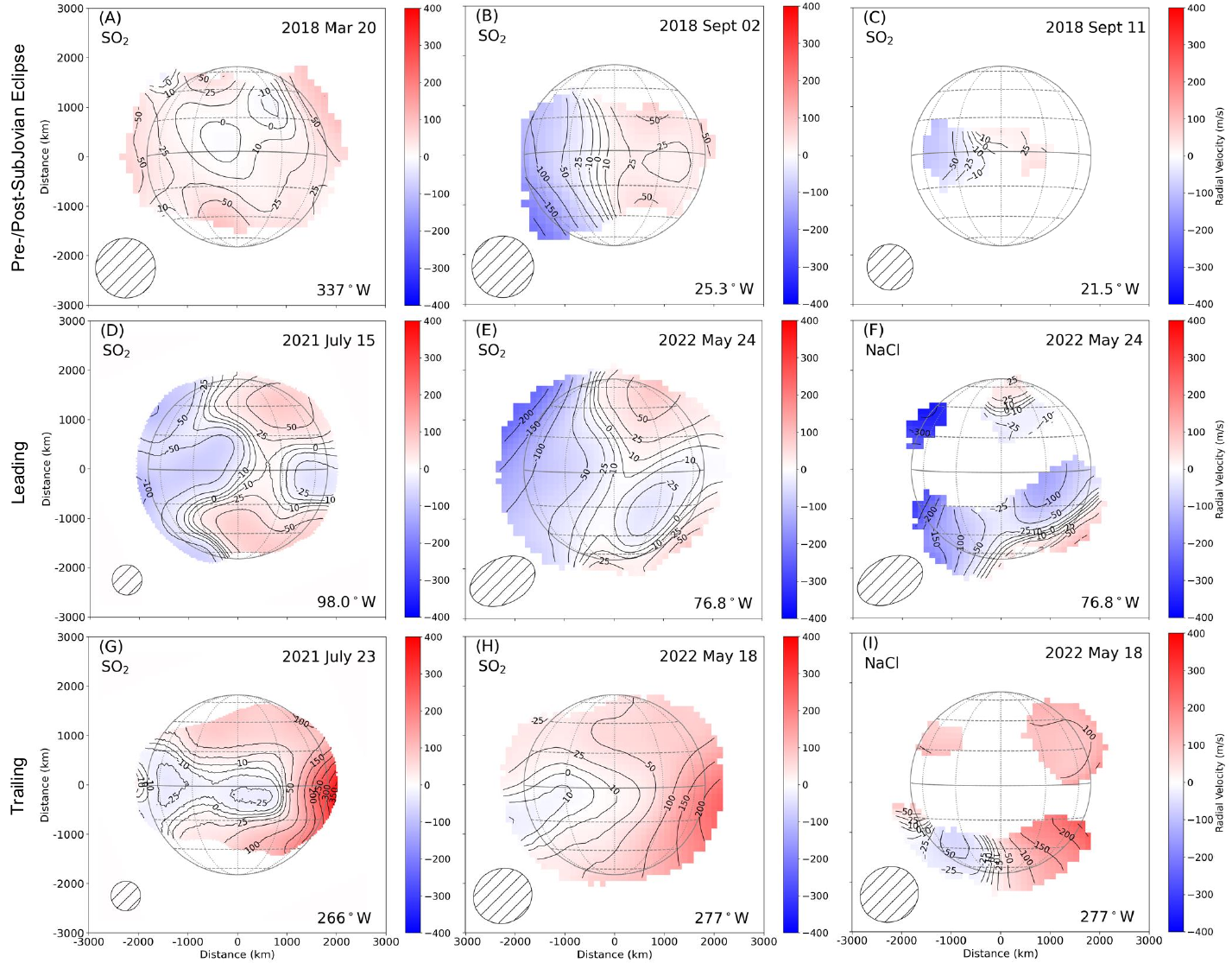}
  \caption{Retrieved SO$_2$ and NaCl LOS velocity maps for pixels with spectral
    SNR$\geq10$ used to derive Io's wind field shown in Figure
    \ref{fig:winds}. The colorbar corresponds to approaching (blue)
    and receding (red) LOS velocities. SO$_2$ velocities on Io's
    subJovian hemisphere pre- and post-eclipse from observations in
    2018 are shown in panels A--C; SO$_2$ velocities at
    Jupiter's eastern and western elongations are shown in panels D
    and G for observations in 2021, E and H for observations in 2022; NaCl velocities from both Jovian
    elongations observed in 2022 are shown in panels F and I. Each
    panel corresponds to those in Figure \ref{fig:winds} before the
    subtraction of Io's solid-body rotation. Io's solid-body radius,
    latitude and longitude lines are denoted by the gray grid, with latitude
    and longitude contours increasing in increments of $22.5^\circ$ and
    $30^\circ$, respectively. The sub-observer central longitude is
    shown in the lower right of each panel, and the ALMA beam (FWHM of the
    PSF) for each observation is shown
  as a hashed ellipse in the lower left of each panel.}
  \label{fig:vels}
\end{figure}

\begin{figure}
  \centering
  \includegraphics[width=0.98\textwidth]{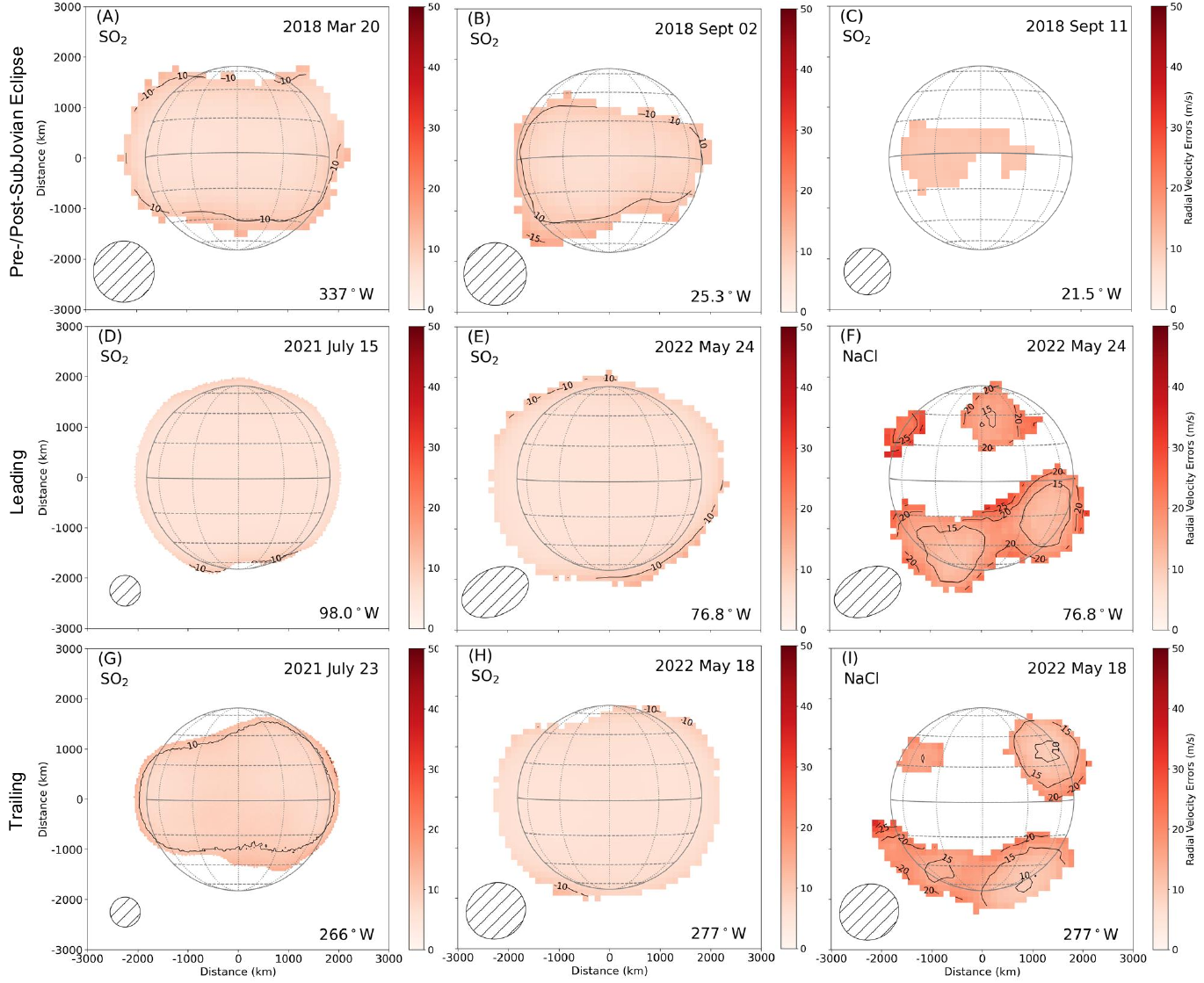}
  \caption{Retrieved velocity error maps corresponding to the velocity
    maps shown in Figure \ref{fig:vels}. Velocity errors are shown where the gas emission is
    $\geq10\times$ the spectral RMS noise and within an ALMA 
    beam-width (shown in the lower left) of Io's surface. Io's solid-body radius,
    latitude ($22.5^\circ$ increments) and longitude ($30^\circ$
    increments) lines are represented by the grid in gray.}
  \label{fig:errs}
\end{figure}

Figure \ref{fig:vels} shows the derived LOS velocities for the
combined SO$_2$ spectral line measurements (panels A--E, G, H) and
NaCl ($J=33-32$) spectral line (panels F, I). The associated errors,
as derived from Monte Carlo estimation, are shown
in Figure \ref{fig:errs}. Following the removal of Io's solid-body
rotation (Appendix \ref{app:rot}), these radial velocity maps were
used to produce the final
wind speeds shown in Figure \ref{fig:winds} in Section \ref{sec:dis}.

These maps show
SO$_2$ LOS radial velocities that vary spatially and fluctuate between
$\sim$\abs{50} m s$^{-1}$ on the disk, reaching maximum
values of 200--300 m s$^{-1}$ towards the limbs, with typical errors
  on order 5--10 m s$^{-1}$. The velocity uncertainties increase
to $\sim15$--20 m s$^{-1}$ as the spectral SNR
degrades, typically at high latitudes and off
disk where the atmosphere is increasingly rarefied as found by
previous observations \citep{de_pater_21, de_pater_23}. The NaCl velocities
consistently reach \textgreater\abs{100} m s$^{-1}$, up to
$\sim$\abs{350} m s$^{-1}$, though with larger associated errors
($\sim15$--30 m s$^{-1}$). The NaCl emission is confined to localized, high column
density regions attributed to
volcanic outgassing \citep{redwing_22, de_kleer_24}, with considerable uncertainty elsewhere. As such,
the spatial distribution of both SO$_2$ and NaCl constrains the
sampling of the atmospheric wind field, in addition to the
aforementioned spectral SNR. While regions of low column
density can still experience high wind velocities (see \citealp{walker_12,
  walker_phd}), these regions primarily occur on the terminators or
nightside, or are difficult to sound with
spectroscopic techniques.

\section{Subtraction of the Underlying Solid-Body Rotation} \label{app:rot}
The radial velocity measurements presented in Appendix \ref{app:maps}
contain a superposition of Io's innate solid-body rotation and any
additional atmospheric motion -- related to sublimation-driven winds,
toroidal momentum transfer, volcanic outgassing, or otherwise. Io's solid-body
angular velocity ($\Omega$) is \abs{75} m s$^{-1}$,
which is significant enough to impact the gas velocities predicted
through models \citep{gratiy_10, walker_12}; this motion may
counteract (and reverse) sublimation-driven flow on the dawn terminator, while reinforcing it
on the dusk terminator. To remove this
effect and isolate velocities due to winds (and their underlying
drivers), we calculated the latitude ($\theta$) and longitude ($\phi$)
of every $x, y$ pixel pair in each ALMA image and determined the modeled solid-body
velocity, $V_{m}(x,y)$, through the relation with the angular velocity:

\begin{equation}
  V_{m}(x,y) = \Omega cos\theta(x,y) sin(\phi(x,y)-\phi_{obs}) cos\theta_{obs}
\end{equation}

where $\theta_{obs}$ and $\phi_{obs}$ correspond to the sub-observer
latitude and longitude, respectively, of each observation (see Table \ref{tab:obs}).

\begin{figure}
  \centering
  \includegraphics[width=0.98\textwidth]{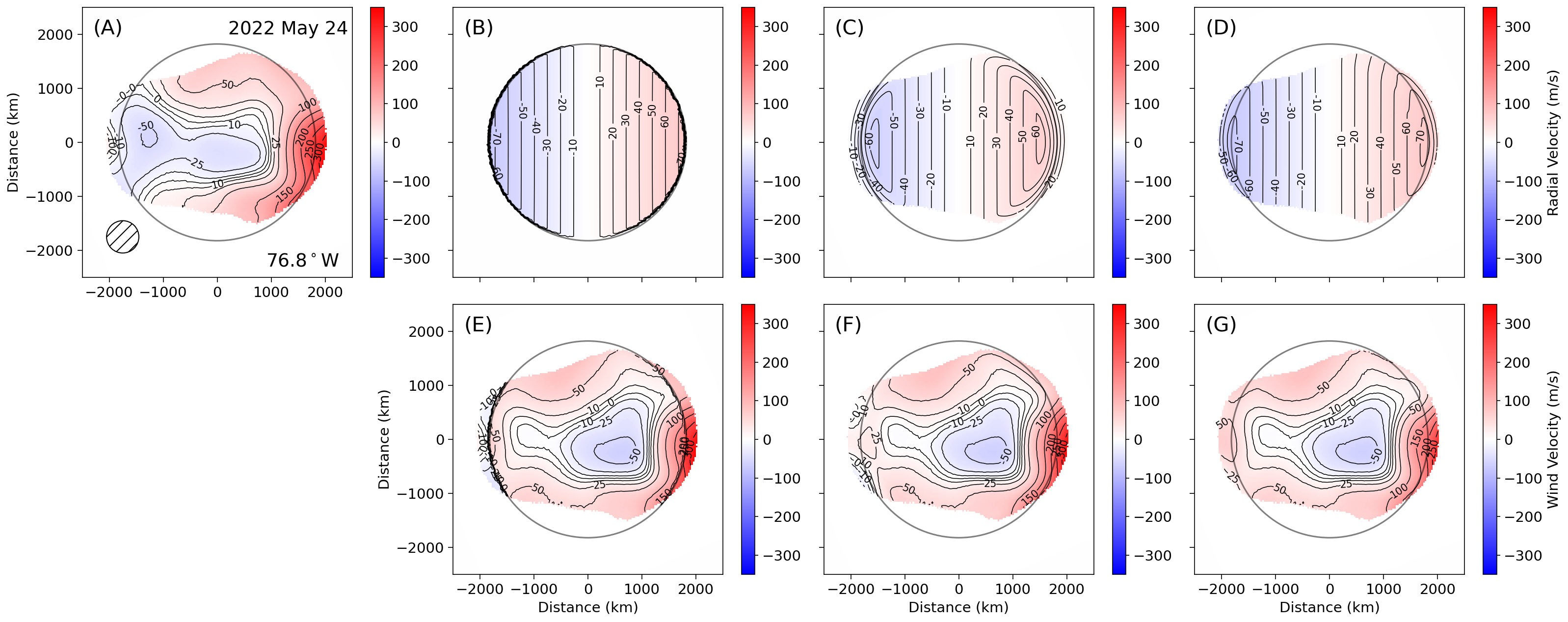}
  \caption{Comparison of the retrieved wind map of the SO$_2$
    ($J_{K_a,K_c}=19_{1,19}-18_{0,18}$) spectral line at 346.7 GHz (A) with
    solid-body rotation models (B--D) and the final, derived wind maps
    (E--G) following the subtraction of the models from the
    data. Different rotation models were tested: a pure solid-body
    rotation model (B);
    the solid-body rotation model convolved with the ALMA beam (C); solid-body
    rotation extending off the surface of Io to the ALMA beam width,
    and subsequently convolved with the beam shape (D).
    The wind fields (E--G) result from the velocity field (A) subtracted by
    the corresponding model in the panel above (e.g., velocities A - model B = winds E).}
  \label{fig:app_rot}
\end{figure}

Figure \ref{fig:app_rot} shows an example of a single retrieved SO$_2$
velocity field (panel A) compared to the model of Io's solid-body rotation as a
function of latitude and longitude (panel B). Without further
modification, the wind field produced by subtracting the
solid-body model
from the measured velocity field is shown in Figure \ref{fig:app_rot}E. However, the smearing of 
SO$_2$ emission across the ALMA beam shape must also be accounted for; the discontinuous winds
exhibited at the limbs in Figure \ref{fig:app_rot}E are caused by subtracting the pure
solid-body velocity from the retrieved velocities, which are innately
convolved by the ALMA beam. Figure \ref{fig:app_rot}C shows the
solid-body model (panel B) convolved with the ALMA beam, a 2D Gaussian
function. While the resulting winds (Figure \ref{fig:app_rot}F) present smoother
gradients at the limbs, the limb velocities themselves are not 
properly accounted for due to the convolution of the ALMA beam shape
with surrounding space at 0 m s$^{-1}$. This is corrected in Figure
\ref{fig:app_rot}D, where the solid-body rotation is projected beyond
Io's radius by an ALMA beam width or to the point at which the
spectral SNR$\leq10$, and subsequently convolved with the beam
shape. Though a somewhat coarse approximation, the connection between
the motion of each atmospheric species and the surface -- before
invoking atmospheric winds --  is difficult to assume based on
the generation of SO$_2$ or NaCl through sublimation, sputtering, or
outgassing, which each may produce gas with a natural velocity
dispersion that is disconnected with the solid-body. However, the
model sufficiently removes edge effects produced through the
subtraction of the solid-body model (Figure \ref{fig:app_rot}B) alone, and the
differences due to the extension of the model into the atmosphere are not
significant compared to the noise towards the limb (compare the limb
velocities in Figure \ref{fig:app_rot}F and G). A more rigorous consideration of these
effects, and the inclusion of an additional component simulating the
torus momentum transfer, should be incorporated into future work. 

\end{appendices}

\pagebreak

\end{document}